\documentclass[pre,twocolumn,showpacs,longbibliography]{revtex4-1}
\usepackage{amsmath,amssymb}
\usepackage{graphicx,color}
\usepackage{times}
\usepackage[pdftex,bookmarks,colorlinks,breaklinks]{hyperref}
\makeatletter
\hypersetup{linkcolor=blue,citecolor=blue,filecolor=dullmagenta,urlcolor=blue, 
 pdftitle={Envelope and phase distribution of a resonance transmission through a complex environment},
 pdfauthor={Dmitry V. Savin}, pdflang={English} }

\newcommand{\aver}[1]{\left \langle #1 \right \rangle}

\newcommand{\Hbg}{H_{\mathrm{bg}}}
\newcommand{\Szero}{S^{(0)}}
\makeatother
\begin{document}

\title{Envelope and phase distribution of a resonance transmission through a complex environment}
\author{Dmitry V. Savin}
 \affiliation{Department of Mathematics, Brunel University London, Uxbridge, UB8 3PH, United Kingdom}
\published{5 June 2018 in \underline{{Phys. Rev. E \textbf{97}, 062202 (2018)}}}

\begin{abstract}
A transmission amplitude is considered for quantum or wave transport mediated by a single resonance coupled to the background of many chaotic states. Such a model provides a useful approach to quantify fluctuations in an established signal induced by a complex environment. Applying random matrix theory to the problem, we derive an exact result for the joint distribution of the transmission intensity (envelope) and the transmission phase at arbitrary coupling to the background with finite absorption. The intensity and phase are distributed within a certain region, revealing essential correlations even at strong absorption. In the latter limit, we obtain a simple asymptotic expression that provides a uniformly good approximation of the exact distribution within its whole support, thus going beyond the Rician distribution often used for such purposes. Exact results are also derived for the marginal distribution of the phase, including its limiting forms at weak and strong absorption.
\end{abstract}
\maketitle

\section{Introduction}\label{sec:intro}

Resonance phenomena have their historical roots in acoustics and mechanical vibrations, with numerous applications ranging from electromagnetism and optics to quantum mechanics and particle physics \cite{Mahaux,Nussenzveig,Mello}. Many of their key features can be understood within a simple model involving scattering on a single resonance (e.g., see \cite{savi16}). Neglecting global phases and assuming invariance under time reversal, the scattering amplitude between any two open channels (say, $c$ and $c'$) is then described by a multichannel Breit-Wigner formula \cite{Mahaux}
\begin{equation}\label{S0_ab}
  \Szero_{cc'}(E) = \delta_{cc'} - i \frac{A_c A_{c'} }{ E-\varepsilon_0+\frac{i}{2}\Gamma_0}\,.
\end{equation}
Here $\varepsilon_0$ is the resonance energy, whereas $A^2_{c}$ and $\Gamma_0$ are the partial (per channel) and total decay widths, respectively. The latter is given by the sum $\Gamma_0=\sum_{c}A^2_c$, with index $c$ running over all channels open at the given scattering energy $E$. This ensures the unitarity of the $S$ matrix (at real $E$).

In many situations, including examples from atomic and nuclear physics \cite{soko97,harn86,soko97i,gu99,zele16} and open mesoscopic systems \cite{aber08,guhr09,soko10,mora12}, such a resonance state is embedded in the background of many complicated states. It represents a specific ``simple'' (deterministic) excitation \cite{soko97}, the amplitude of which spreads over the chaotic states. The arising competition between two damping mechanisms, escape to the continuum and spreading over the background, results in the suppression of transmission through such a simple mode when coupling to the background is increased \cite{savi17}. In such a context, the model provides a useful approach for quantifying fluctuations in an established transmission signal induced by a complex environment, with the exact results being recently presented for the intensity distribution  \cite{savi17} as well as for the joint distribution of the reflection and transmission \cite{savi17a}.

In this work, we explore this direction further and develop a systematic approach for studying joint statistics of the intensity (envelope) and phase of the transmission amplitude. Such a problem is important, e.g., for fading and related applications in wireless communications \cite{Simon,yeh12}. We derive exact nonperturbative results for the joint (intensity-phase) probability density function (jpdf) as well as for the marginal pdf of the transmission phase. The obtained results are valid at any coupling strength to the background and arbitrary absorption that is typically present in real environments. We also discuss in detail the limiting forms of the distributions in the physically interesting cases of weak and strong absorption.

\section{Statistical model}\label{sec:model}
%
We follow Refs.~\cite{savi17,savi17a} to account for the influence of the chaotic background with finite dissipative losses on the resonance scattering. Within the well-known strength function formalism \cite{soko97,Bohr}, such an interaction results in the modified energy dependence that is determined by the substitution $[E-\varepsilon_0+\frac{i}{2}\Gamma_0]^{-1}\to[E-\varepsilon_0+\frac{i}{2}\Gamma_0-g(E)]^{-1}$ in (\ref{S0_ab}). The strength function $g(E)\equiv V^{\dag}(E-\Hbg)^{-1}V$ describes the coupling to the background Hamiltonian $\Hbg$, with $V$ being a vector of the (real) coupling amplitudes. The $S$ matrix can then be represented in the following convenient form:
\begin{equation}\label{S_ab}
  S(E) - 1 = t(E)\,[S^{(0)}(E) - 1]\,,
\end{equation}
where $\Szero$ stands for the $S$ matrix without the background, see Eq.~(\ref{S0_ab}), and the (scalar) function $t(E)$ is defined as follows
\begin{equation}\label{t(E)}
 t(E) = \frac{1 + 2i(\varepsilon_0-E)/\Gamma_0}{1 + 2i[\varepsilon_0-E +g(E)]/\Gamma_0}\,.
\end{equation}
By construction, $t(E)$ has the meaning of the transmission amplitude rescaled in units of that in the ``clean'' system.

\begin{figure*}
  \centering
  \includegraphics[width=.975\linewidth]{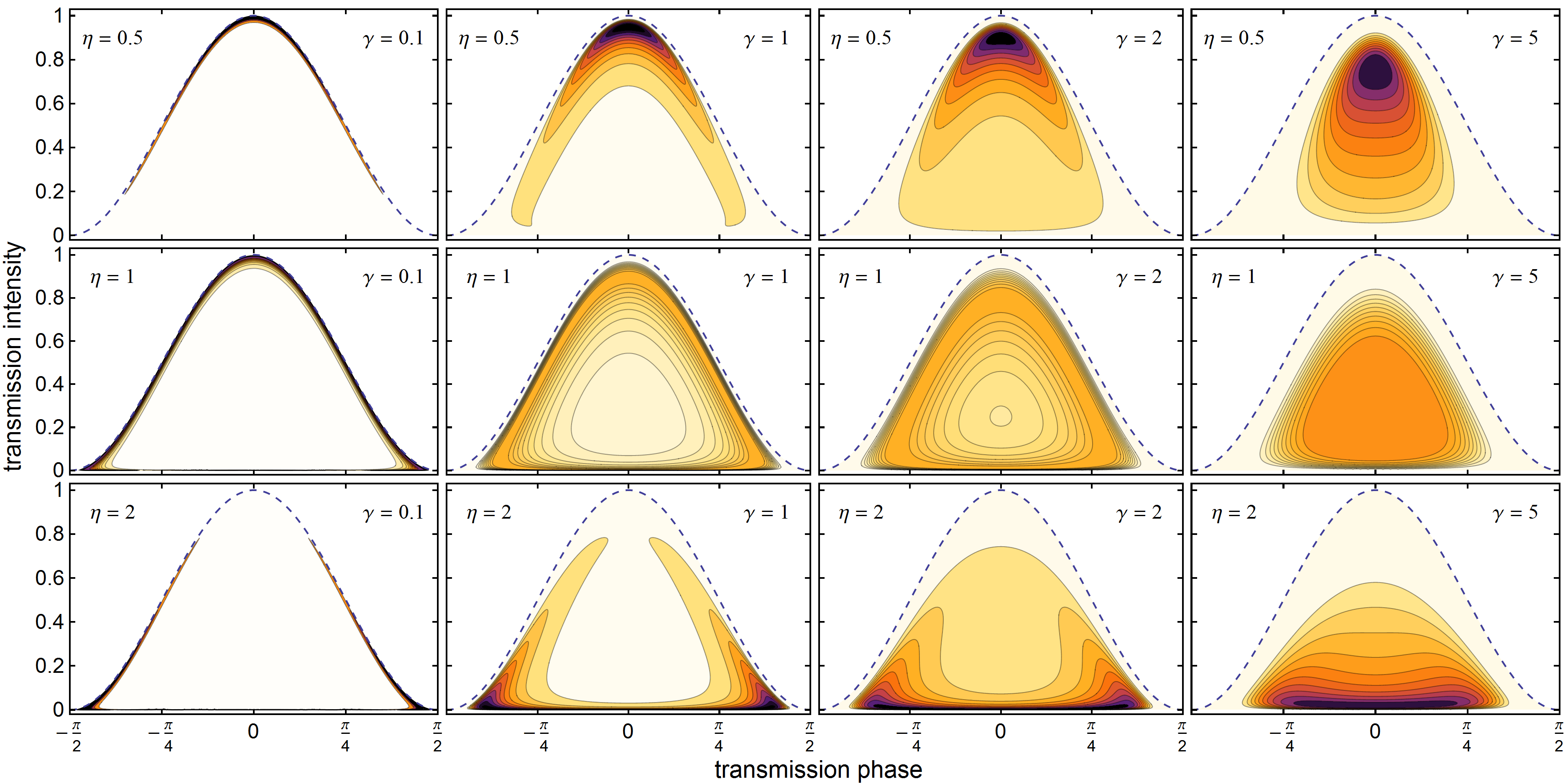}
  \caption{Contour plots of the joint distribution (\ref{Pjoint}) of the transmission intensity ($T$) and phase ($\theta$) at the background coupling $\eta=0.5,1,2$ (rows) and absorption rates $\gamma = 0.1,1,2,5$ (columns). Darker regions correspond to higher values of the jpdf, with black standing for the values of $40,5,3.5,3$ at $\gamma = 0.1,1,2,5$, respectively. The dashed line indicates the boundary $T=\cos^2\theta$ of the distribution support.
  \label{fig:jpdf} }
\end{figure*}

The background typically has a very dense spectrum involving very many, $N\gg1$, highly complicated (chaotic) states. This gives rise to wild fluctuations of $t(E)$ on the local scale of the mean level spacing $\Delta\sim1/N$. We are interested in those at the resonance energy $\varepsilon_0$, corresponding to the peak of the original signal.  When averaged over this fine structure, the scattering amplitudes acquire an additional effective damping $\Gamma_{\downarrow}\equiv2\mathrm{Im}\langle{g(\varepsilon_0-i0 )}\rangle = 2\pi\|V\|^2/(N\Delta)$, the so-called spreading width \cite{soko97,Bohr}. This sets a second decay energy scale in addition to the escape width $\Gamma_0$. The ratio
\begin{equation}
  \eta=\Gamma_{\downarrow}/\Gamma_0
\end{equation}
serves as a natural parameter that describes the competition between the two decay mechanisms, thus controlling the background coupling strength \cite{savi17}. The resonant transmission amplitude $t\equiv t(\varepsilon_0)$ can then be represented as follows:
\begin{equation}\label{t}
  t = (1+i\eta K)^{-1} \equiv\sqrt{T} e^{i\theta}\,,
\end{equation}
where $K=2g(\varepsilon_0)/\Gamma_{\downarrow}$. By construction, $K$ has the meaning of the local Green's function of the background \cite{fyod04b}.

Aiming to describe the universal statistics of both the transmission intensity $T$ and phase $\theta$, we adopt the standard route \cite{guhr98,Stoeckmann,fyod11ox} and model $\Hbg$ by a random $N \times N$ matrix drawn from the Gaussian orthogonal ensemble (GOE). The established connection to the background spectrum enables us also to take into account finite absorption by introducing uniform broadening $\Gamma_\mathrm{abs}$ of the background states. Since such a damping is operationally equivalent to the purely imaginary energy shift, $K\equiv\frac{2}{\Gamma_{\downarrow}} g(\varepsilon_0+\frac{i}{2}\Gamma_\mathrm{abs})$ becomes complex \cite{fyod04b},
\begin{equation}\label{K}
  K =  u - iv, \quad v>0,
\end{equation}
and normalized as $\aver{K}=-i$. The mutually correlated random variables $u$ and $v$ are known to have universal statistics described by the following jpdf \cite{fyod04b}:
\begin{align} \label{P(u,v)}
  P(u,v) = \frac{1}{2\pi v^2}P_0(x), \quad x=\frac{u^2+v^2+1}{2v}>1.
\end{align}
The function $P_0(x)$ is solely determined by the (dimensionless) absorption rate $\gamma \equiv 2\pi \Gamma_\mathrm{abs}/\Delta$  and is known exactly \cite{savi05,fyod05}. In particular, it has an important symmetry, being invariant under the interchange $iK\to1/iK$. This results in
\begin{equation}\label{<t>}
  \aver{t}=(1+\eta)^{-1}
\end{equation}
for the mean value of the transmission amplitude at any $\gamma$.

\section{Joint distribution of $T$ and $\theta$}\label{sec:jpfd}
%
The intensity $T = |t|^2$ and phase $\theta=\arg(t)$ of the transmission amplitude (\ref{t}) are readily found  as follows
\begin{subequations}\label{T,phase}
 \begin{align}
  T & = \frac{1}{(1 + \eta v)^2+\eta^2u^2} \qquad\  (0\leq T \leq 1),
  \\
  \theta & = -\arctan\frac{\eta u}{1+\eta v} \qquad \textstyle
   (-\frac{\pi}{2}\leq\theta\leq\frac{\pi}{2})\,.
 \end{align}
\end{subequations}
We will now derive and study the joint distribution $\mathcal{P}_{\eta,\gamma}(T,\theta)$ of these two quantities at arbitrary values of $\eta$ and $\gamma$.

Let us first consider the case of zero absorption. Then $K=u$ is real, where the random variable $u$ is known to have the standard Cauchy distribution \cite{fyod04b,mell95}. The joint distribution of $T$ and $\theta$ is found by an easy integration and reads
\begin{align}
  \mathcal{P}_{\eta,\gamma=0}(T,\theta)
  &= \int\limits_{-\infty}^{\infty}\frac{du}{\pi}
     \frac{\delta(\tan\theta+\eta u)}{1+u^2} \delta\Bigl(T - \frac{1}{1 + \eta^2u^2}\Bigr) \nonumber
  \\
  \label{Pjoint0}
  &=  \delta(T - \cos^2\theta) \mathcal{P}_0(\theta).
\end{align}
The first (singular) factor here represents the conditional pdf of $T$ and $\mathcal{P}_0(\theta)$ stands for the marginal pdf of $\theta$ given by
\begin{equation}\label{Pphase0}
  \mathcal{P}_0(\theta) = \frac{1}{\pi (\eta\cos^2\theta+\eta^{-1}\sin^2\theta)}.
\end{equation}
The phase distribution (\ref{Pphase0}) has a local maximum (minimum) at $\theta=0$ for $\eta<1$ ($\eta>1$) and becomes uniform at the special coupling $\eta=1$. Note that this function has the following symmetry $\mathcal{P}_0(\theta)|_\eta=\mathcal{P}_0(\theta\pm\frac{\pi}{2})|_{\eta^{-1}}$ under the involution $\eta\to\eta^{-1}$. This can be traced back to the more general symmetry property satisfied by the joint distribution of transmission and reflection at arbitrary absorption established in \cite{savi17a}.

In the general case of finite absorption, the singularity of the joint distribution is removed, since $T$ and $\theta$ are no longer functions of each other. The jpdf in question is then obtained from the known function $P(u,v)$ by making use of relations (\ref{T,phase}). To this end, it is convenient first to choose $y=\tan\theta$ instead of $\theta$, so $|\partial y/\partial\theta|=1+y^2$, and then to compute the Jacobian $|\partial(u,v)/\partial(T,y)|=1/[2\eta^2T^2(1+y^2)]$. Noticing further that $v=\frac{1}{\eta}(|\cos\theta|/\sqrt{T}-1)\geq0$ and collecting all factors together, we arrive at the following exact result:
\begin{equation}\label{Pjoint}
  \mathcal{P}_{\eta, \gamma}(T,\theta) =
  \frac{\Theta(\cos\theta-\sqrt{T})}{4\pi T(\cos\theta-\sqrt{T})^2}
  P_0[\mathrm{x}_\eta(T,\theta)],
\end{equation}
where $\Theta(x)$ is the Heaviside step function and
\begin{equation}\label{x}
 \mathrm{x}_\eta(T,\theta) =
 \frac{T(1+\eta^2)-2\sqrt{T}\cos\theta+1}{2\eta\sqrt{T}(\cos\theta-\sqrt{T})}.
\end{equation}
This clearly shows that the transmission intensity and phase exhibit strong statistical correlations at finite absorption.

The joint distribution (\ref{Pjoint}) is nonzero for $0\leq T \leq\cos^2\theta$. Its profile within this region is controlled by two parameters $\gamma$ and $\eta$. In the limit of vanishing absorption, $\gamma\to0$, one can use \cite{fyod04b} that $P_0(x)\to\delta(\frac{1}{x})$ which readily yields the zero absorption result (\ref{Pjoint0}). At finite absorption, the function $P_0(x)$ gets exponentially suppressed ($\sim e^{-\gamma x/4}$) for large $x\gg1$. As a result, the distribution at small $\gamma$ is mostly concentrated within a thin layer $\sim\gamma\ll1$ near the boundary $T=\cos^2\theta$.  When $\gamma$ is increased, the distribution starts exploring its whole support. Its weight is gradually moved from the central region around $T\sim1-2\eta$ at $\eta\ll1$ to a stripe around $T\sim\eta^{-2}$ at $\eta\gg1$. All these features are clearly seen in Fig.~\ref{fig:jpdf} showing the density plots of $\mathcal{P}_{\eta,\gamma}(T,\theta)$ for various values of $\eta$ and $\gamma$.

It is worth discussing the statistical correlations between $T$ and $\theta$ in more detail. It is natural to expect that such correlations should go away, when absorption becomes very strong. In the limit $\gamma\gg1$, it is actually known \cite{fyod04b} that the function $P_0(x)$ simplifies to $P_0(x)\approx\frac{\gamma}{4}\,e^{-\frac{\gamma}{4}(x-1)}$. With this limiting form, we readily get the following asymptotic expression:
\begin{equation}\label{Pasym}
  \mathcal{P}^{\mathrm{(asym)}}_{\eta, \gamma\gg1}(T,\theta) =
  \frac{ \gamma\exp\bigl[ -\frac{\gamma(1+\eta)^2}{ 8\eta }
        \frac{T-2\aver{t}\sqrt{T}\cos\theta +\aver{t}^2 }{ \sqrt{T}(\cos\theta-\sqrt{T})}\bigr]
  }{ 16\pi T(\cos\theta-\sqrt{T})^2}\,,
\end{equation}
where $\aver{t}$ is given by (\ref{<t>}) and $0\leq T \leq\cos^2\theta$. Therefore, the correlations remain essential even at strong absorption.

Still, if we were to assume uncorrelated $T$ and $\theta$, it would imply that $\aver{\sqrt{T}}=\aver{t}$ and $\aver{\theta}=0$. One can then perform an expansion over the fluctuations in (\ref{Pasym}), keeping up to the quadratic terms in the exponential there and replacing $T$ and $\theta$ with their mean values elsewhere. This results in the jpdf being factorized as a product of two Gaussian distributions,
\begin{equation}\label{Pnorm}
  \mathcal{P}^{\mathrm{(gaus)}}_{\eta, \gamma\gg1}(T,\theta) =
  \frac{\exp\bigl[ -\frac{1}{2\sigma^2_T}(\sqrt{T}-\aver{t})^2
  -\frac{\theta^2}{2\sigma^2_\theta}\bigr] }{ 4\pi\sigma_T\sigma_\theta\aver{t}}\,,
\end{equation}
where the corresponding variances are given by
\begin{equation}\label{var}
  \sigma^2_T = \frac{4\eta^2}{\gamma(1+\eta)^4}, \quad \sigma^2_\theta=\frac{4\eta^2}{\gamma(1+\eta)^2}.
\end{equation}
The overall factor $\frac{1}{2\aver{t}}$ takes automatically into account the proper normalization. (It appears since it is the variable $\sqrt{T}$, not $T$, that becomes Gaussian distributed at very large $\gamma\gg1$.) We note, however, that the Gaussian approximation (\ref{Pnorm}) is rather crude, because of the finite support of the exact distribution (\ref{Pjoint}). However, one can obtain a better approximation at strong absorption by studying the statistics of the real and imaginary parts of $t$ instead. We will do that now.

\begin{figure}
  \centering
  \includegraphics[width=.975\linewidth]{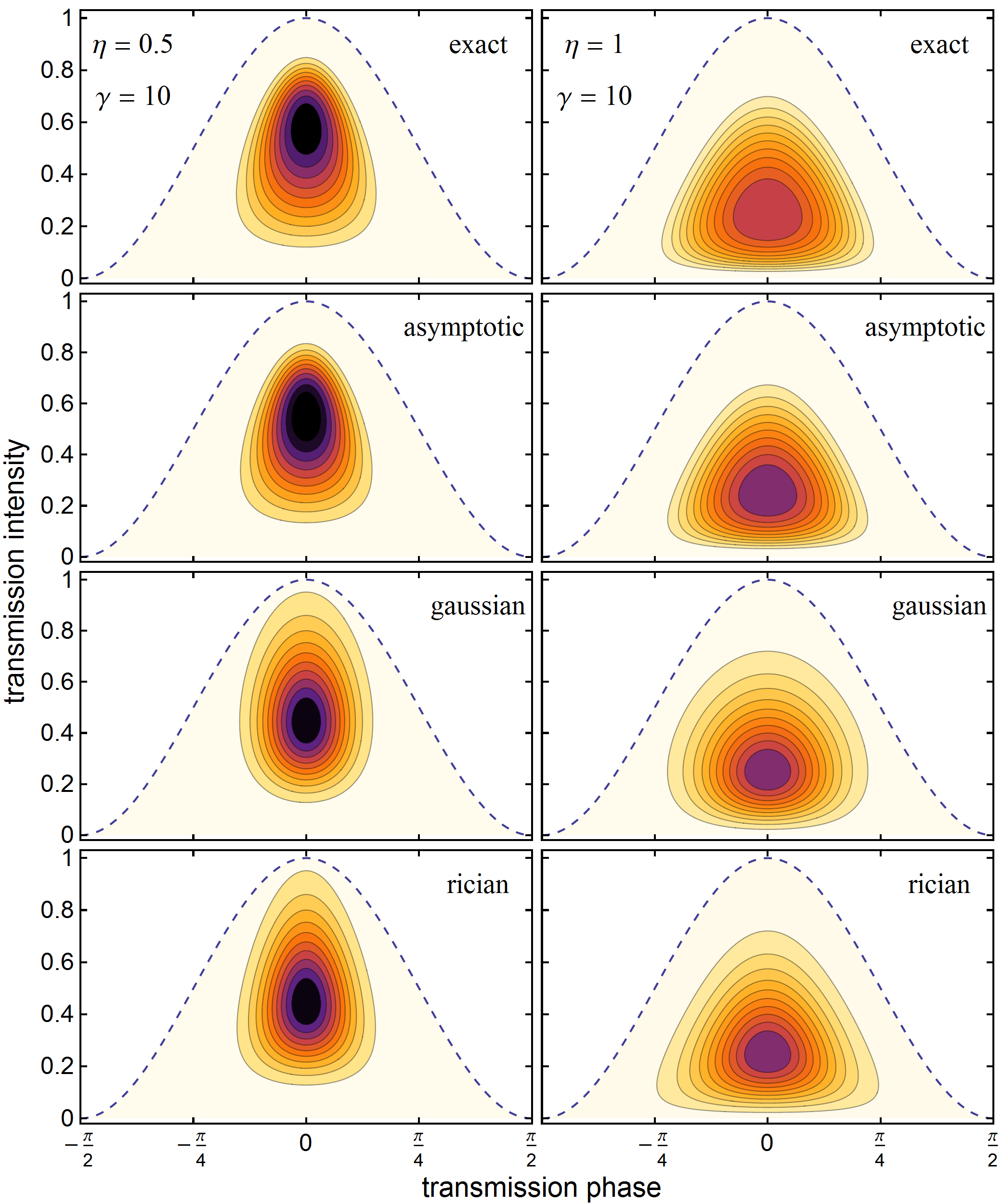}
  \caption{Comparison between the exact jpdf (\ref{Pjoint}), its asymptotic form (\ref{Pasym}) and the Gaussian (\ref{Pnorm}) and Rician (\ref{Price}) approximations at strong absorption. The background coupling $\eta=0.5$ (left panel) or $1$ (right panel) and the absorption rate $\gamma=10$ everywhere.
  \label{fig:jappr} }
\end{figure}

Let $t=t_r+it_i$. Clearly, we have $\aver{t_r}=\aver{t}$ and $\aver{t_i}=0$ for the mean values. The joint distribution of $t_r$ and $t_i$ can be derived along the same lines as presented above. Expressing $t_r,t_i$ in terms of $u,v$ and computing the corresponding Jacobian, we finally get the following exact jpdf (at any $\gamma$):
\begin{align}\label{PjontRI}
  \mathcal{P}_{\eta, \gamma}(t_r,t_i) =
  \frac{ P_0\bigl(1+\frac{(1+\eta)^2}{2\eta s}[(t_r-\aver{t})^2+t_i^2]\bigr) }{ 2\pi\eta s^2}
\end{align}
within the semicircle region $s\equiv t_r(1-t_r)-t_i^2>0$ and zero otherwise. Expression (\ref{PjontRI}) clearly reveals significant correlations between the real and imaginary parts of the transmission amplitude as well. However, $t_r$ and $t_i$ appear to get uncorrelated faster than $T$ and $\theta$ when absorption grows. Indeed, performing the fluctuation analysis of (\ref{PjontRI})  at $\gamma\gg1$, one finds that both $t_r-\aver{t}$ and $t_i$ become uncorrelated normal variables with the same variance $\sigma^2_T$ given by (\ref{var}). In such an approximation, finding the amplitude and phase distributions of $t_r=\sqrt{T}\cos\theta$ and $t_i=\sqrt{T}\sin\theta$ reduces to a classical problem studied by Rice~\cite{rice48} (see also \cite{yaco05}), yielding
\begin{equation}\label{Price}
  \mathcal{P}^{\mathrm{(rice)}}_{\eta, \gamma\gg1}(T,\theta) =
  \frac{1}{ 4\pi\sigma^2_T}e^{-(T-2\aver{t}\sqrt{T}\cos\theta+\aver{t}^2)/2\sigma^2_T} \,.
\end{equation}
The Rician approximation (\ref{Price}) resembles the exact asymptotic form (\ref{Pasym}) in its structure, but fails to properly take into account the boundaries of the distribution support. For that reason, it provides a reasonable approximation only at $\eta\approx1$, when the density is mostly concentrated in the centre, showing noticeable deviations otherwise, when the density gets concentrated near $T\sim1$ ($T\sim\eta^{-2}$) for small (large) $\eta$. Note that our asymptotic result (\ref{Pasym}) is free from such shortcomings, providing good uniform approximation even at moderately large $\gamma$. This discussion is further illustrated in  Fig.~\ref{fig:jappr}.

With the exact result (\ref{Pjoint}) in hand, one can further obtain both marginal and conditional pdf's by performing the relevant integrations. In particular, the distribution of the transmission intensity can be brought to the form recently derived and thoroughly studied in \cite{savi17}. Therefore, we will concentrate only on the analysis of the phase distribution below.

\section{Phase distribution}\label{sec:phase}
%
The distribution $\mathcal{P}(\theta)$ of the transmission phase is obtained by integrating (\ref{Pjoint}) over $T$. Setting $T=\cos^2\theta/(1+p)^2$ and choosing $p>0$ as a new integration variable, we arrive after some algebra at the following convenient representation:
\begin{equation}\label{Pphase}
  \mathcal{P}(\theta) =  \frac{\sec^2\theta}{2\pi} \int_{0}^{\infty}\frac{dp}{p^2}
   (1+p)P_0 [x(p,\theta)],
\end{equation}
where $\sec\theta=(\cos\theta)^{-1}$ and $x(p,\theta)$ is defined as follows
\begin{equation}\label{x}
  x(p,\theta) =\frac{(1+p)^2\sec^2\theta - 2p + \eta^2 - 1}{2\eta p}\,.
\end{equation}
With an explicit formula for $P_0$ found in \cite{savi05}, Eq.~(\ref{Pphase}) provides the exact solution to the problem at arbitrary $\eta$ and $\gamma$.

\begin{figure}
  \centering
  \includegraphics[width=.925\linewidth]{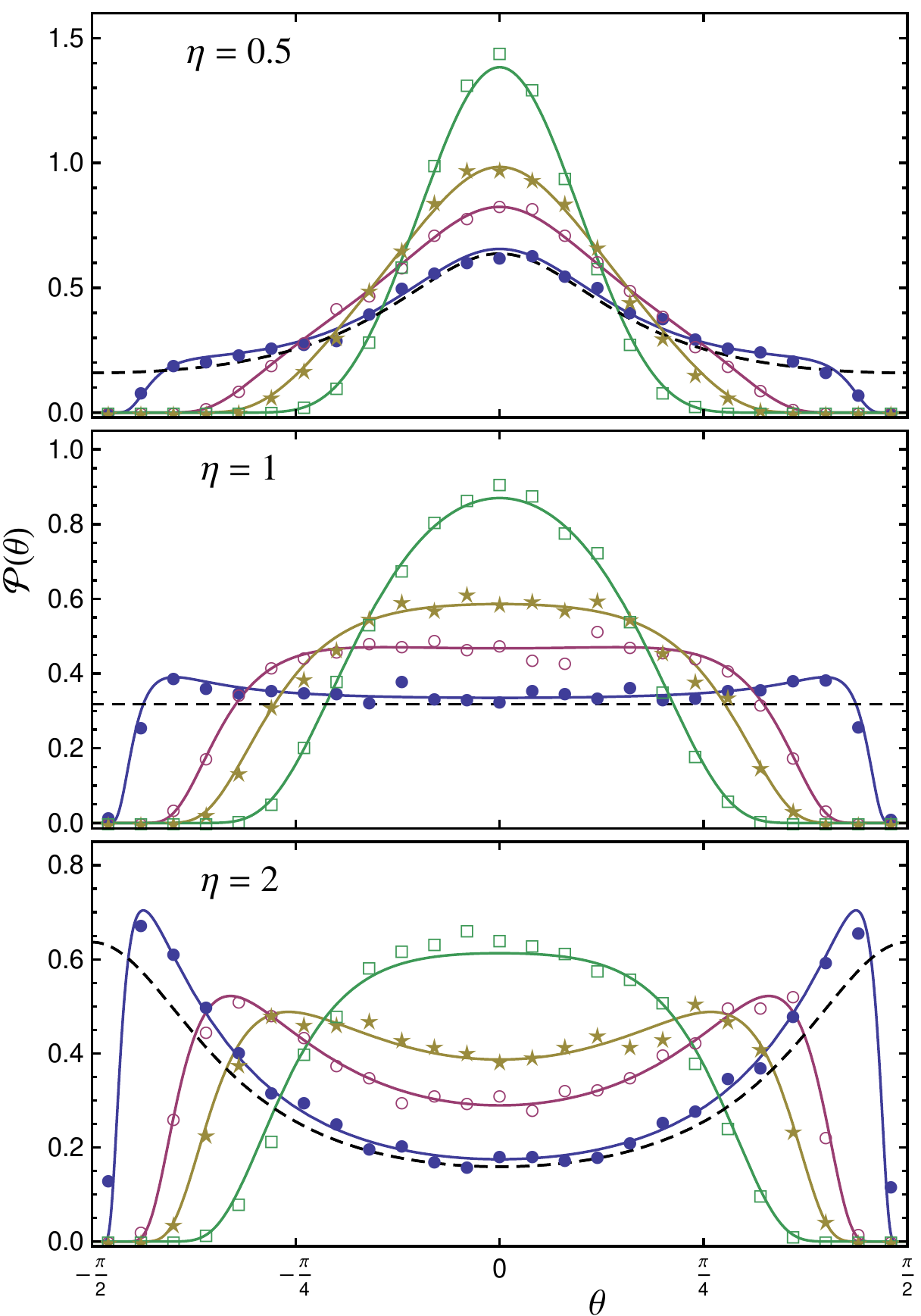}
  \caption{Distribution of the transmission phase for the background coupling $\eta=0.5$ (top),  $1$ (middle) and $2$ (bottom)  and the absorption rate $\gamma=0.1\,(\bullet), 1\,(\circ), 2\,(\star)$ and $5\,(\Box)$. The solid lines show the analytical prediction (\ref{Pphase}), whereas the symbols stand for numerics with $10^4$ realizations of $200{\times}200$ random GOE matrices. The dashed lines show the zero absorption result (\ref{Pphase0}) for comparison.
  \label{fig:phase} }
\end{figure}
Further analysis is possible in the physically interesting limiting cases of weak and strong  absorption, when the function $P_0$ is known to take simpler exact asymptotic forms \cite{fyod04b}. At $\gamma\ll1$, one has $P_0(x)\approx\frac{2}{\sqrt{\pi}}(\frac{\gamma}{4})^{3/2}\sqrt{x+1}\,e^{-\frac{\gamma}{4}(x+1)}$. A close inspection of Eq.~(\ref{Pphase}) shows that the dominant contribution to the integral comes from small $p\ll1$. In the leading order, one can approximate $\frac{\gamma}{4}(x+1)\approx \mu+\rho/p$, where $\mu$ and $\rho$ are certain functions of $\theta$ (see below), and replace $(1+p)$ by unity in the integration measure. Performing the integration, one finds the following approximation at weak absorption:
\begin{equation}\label{Pphase_weak}
  \mathcal{P}_{\gamma\ll1}(\theta)\approx\mathcal{P}_{0}(\theta)
     \Bigl[\mathrm{erfc}(\sqrt{\mu}) + 2\sqrt{\mu/\pi}e^{-\mu}\Bigr],
\end{equation}
where $\mu=\frac{\gamma}{4\eta}(\sec^2\theta-1+\eta)$ and $\mathrm{erfc}(z)=1-\mathrm{erf}(z)$ is the complementary error function. The bulk of distribution (\ref{Pphase_weak}) is essentially given by that at zero absorption, Eq.~(\ref{Pphase0}). The correction factor becomes crucial near the edges, where the exact distribution has an exponential cutoff $\sim e^{-(\gamma/4\eta)\sec^2\theta}$.

In the opposite case of strong absorption, $\gamma\gg1$, one can use that $P_0(x)\approx\frac{\gamma}{4}\,e^{-\frac{\gamma}{4}(x-1)}$. The resulting integration can be done exactly without further assumptions, yielding
\begin{equation}\label{Pphase_str}
  \mathcal{P}_{\gamma\gg1}(\theta)\approx \frac{\gamma\sec^2\theta}{4\pi}
  \Bigl[K_0(\xi) + \frac{\gamma\sec^2\theta}{4\eta\xi} K_1(\xi)\Bigr]e^{-\nu},
\end{equation}
with $\xi=\frac{\gamma}{4\eta}\sec\theta\sqrt{\sec^2\theta-1+\eta^2}$, $\nu=\frac{\gamma}{4\eta}(\sec^2\theta-1-\eta)$, and $K_{n}(z)$ being the modified Bessel function. We note that distribution (\ref{Pphase_str}) is properly normalized to unity at any $\gamma$.  But it approximates the exact one only for large $\gamma$, of course. In the limit of very large $\gamma\gg1$, expression (\ref{Pphase_str}) tends to a Gaussian with zero mean and  the variance $\sigma^2_\theta$ provided in (\ref{var}).

Figure~\ref{fig:phase} illustrates this analysis and shows the phase distribution for the same values of $\eta$ and $\gamma$ as in Fig.~\ref{fig:jpdf}. Also shown are the results of numerical simulations of the transmission phases (\ref{T,phase}) with random GOE matrices. Since the explicit form of the exact function $P_0(x)$ is fairly involved \cite{savi05}, we have used instead a much simpler interpolation formula,
\begin{equation}\label{Pzero_int}
  P_0^{\mathrm{(int)}}(x) = C_\gamma^{-1}
  \bigl(A_{\gamma}\sqrt{\gamma(x+1)} + B_\gamma\bigr) e^{-\frac{\gamma}{4}(x+1)}\,,
\end{equation}
suggested in \cite{fyod04b}, to draw the analytical curves. Here, the $\gamma$ dependent constants $A_\gamma=(e^{\gamma/2}-1)/2$, $B_\gamma=1+\frac{\gamma}{2}-e^{\gamma/2}$, and
$C_\gamma = \frac{4}{\gamma} [2\Gamma(\frac{3}{2},\frac{\gamma}{2})A_\gamma+e^{-\gamma/2}B_\gamma]$,
with $\Gamma(\nu,\alpha)$ being the upper incomplete gamma function. This interpolation formula was previously found to work with very good accuracy when compared to the exact results, including the transmission distribution \cite{savi17}.  The present case is no exception, as is seen from the flawless agreement with the numerics.

\section{Conclusions}\label{sec:summary}
%
In this work, we have presented a systematic study of fluctuations and correlations in the transmission intensity and phase induced by coupling the transmitting resonance to the chaotic background. Modeling such a background by RMT, we have derived the joint (``envelope--phase'') distribution in the exact form (\ref{Pjoint}) valid at arbitrary coupling to and losses in the background. The intensity and phase are found to exhibit strong and non-trivial statistical correlations within certain finite support of the joint distribution. These correlations remain essential even in the limit of strong absorption. In the latter case, we have discussed in detail the relevant approximations and their accuracy. In particular, simple asymptotic expression (\ref{Pasym}) has been obtained that, in contrast to the Gaussian or Rician approximations, provides uniformly good approximation within the whole distribution support.

We have also derived the distribution of the transmission phase and studied its exact limiting forms in the physically interesting cases of weak and strong absorption. The analytical results are supported by straightforward numerical simulations performed with random GOE matrices. We note that recent advances in experimental techniques have provided access to the full $S$ matrix, including the phases, in various microwave cavity experiments ~\cite{kuhl05,kuhl05a,hemm06,koeb10,diet10,kuhl13,grad14}. In particular, exact nonperturbative results for the statistics of diagonal \cite{fyod05,savi05} and off-diagonal \cite{kuma13,nock14} $S$ matrix elements were tested with high accuracy in such experimental studies. The approach developed here depends essentially on two model parameters, the background coupling ($\eta$) and absorption strength ($\gamma$), which can be extracted from scattering data as discussed in \cite{savi17appa}.  Thus, we expect that our results will find further applications within a broader context of wave chaotic systems.

\acknowledgments

I am grateful to the Organizers of the 8th Workshop on Quantum Chaos and Localisation Phenomena in Warsaw, Poland, 19--21 May 2017, for hospitality during my stay in Warsaw. I thank my collaborators of Ref.~\cite{savi17} for useful discussions of relevant experimental studies with a reverberation chamber setup developed at INPHYNI, Nice.

%

\end{document}